\documentclass[12pt]{article}
\usepackage{bbm}
\usepackage{amsfonts}
\usepackage{latexsym,cite,amssymb}
\usepackage{times}
\usepackage{graphicx}
\usepackage{color}
\def\be{\begin{equation}}
\def\ba{\begin{eqnarray}}
\def\ee{\end{equation}}
\def\ea{\end{eqnarray}}

\newcommand{\del}{\partial}

\newcommand{\dd}{{\rm d}}

\makeatletter
\renewcommand{\theequation}{\thesection.\arabic{equation}}
\@addtoreset{equation}{section} \makeatother
\input amssym.def
\input amssym.tex

\begin{document}

\title{\vskip -60pt
\vskip 20pt 
 Holographic RG flows and transport coefficients in
Einstein-Gauss-Bonnet-Maxwell theory~}
\author{ Xian-Hui Ge${}^{1}~$, Yi Ling${}^{2}~$, Yu Tian${}^{3}~$
and  Xiao-Ning Wu${}^{4}~$}
\date{}
\maketitle \vspace{-1.0cm}
\begin{center}
~~~
${}^{1}$Department of Physics, Shanghai University, Shanghai 200444, China\\
~{}
${}^{2}$Institute of High Energy Physics,
Chinese Academy of Sciences, Beijing 100049, China and Center for Relativistic Astrophysics and High Energy Physics,
Department of Physics, Nanchang University, 330031, China \\
~{}
${}^{3}$College of Physical Sciences, Graduate University of
Chinese Academy of Sciences, Beijing 100049, China\\
~{}
${}^{4}$Institute of Mathematics, Academy of Mathematics
and System Science, CAS, Beijing 100190, China and
Hua Loo-Keng Key Laboratory of Mathematics, CAS, Beijing 100190, China\\
{\small {E-mail: {{gexh@shu.edu.cn}, ~{yling@ncu.edu.cn},~{ytian@gucas.ac.cn},~{wuxn@amss.ac.cn}}}}
~~~\\
~~~\\
\end{center}

\begin{abstract}
We apply the  membrane paradigm and the holographic Wilsonian
approach to the  Einstein-Gauss-Bonnet-Maxwell theory. The
transport coefficients for a  quark-gluon plasma living on
the cutoff surface are derived in a spacetime of charged black
brane.  Because of the mixing of the Gauss-Bonnet coupling and
the Maxwell fields, the vector modes/shear modes of the metric and
Maxwell fluctuations turn out to be very difficult to decouple. We
firstly evaluate the AC conductivity at a finite cutoff surface by
solving the equation of motion numerically, then manage to derive
the radial flow of DC conductivity with the use of the Kubo
formula. It turns out that our analytical results match the
numerical data in low frequency limit very well.  The
diffusion constant $D(u_c)$ is also derived in a long wavelength
expansion limit. We find it depends on the Gauss-Bonnet coupling
as well as the position of the cutoff surface.

 \end{abstract}
{\small
\begin{flushleft}
\end{flushleft}}
\newpage
\tableofcontents 

\section{Introduction }

Recently, a holographic version of Wilsonian renormalization group
(hWRG) has been proposed to describe field theories with a finite
cut-off \cite{pol,faulk,strominger,nick}. The essential idea of
hWRG is to integrate out the bulk field from the boundary up to
some intermediate radial distance. The radial direction in the
bulk marks the energy scale of the boundary theory and the radial
flow in the bulk geometry can be interpreted as the
renormalization group flow of the boundary
theory\cite{sus,peet,ak,al,gi,di,ba,free,boer,boer1,li}.
Moreover, some evidence has been presented in \cite{sin}
indicating that the membrane paradigm and the hWRG method are
probably equivalent.

In \cite{strominger} and later in \cite{sz}, it was shown that the
diffusion coefficient for Reissner-Nordstrom-Anti de
Sitter(RN-AdS) black holes can be calculated in the scaling limit
without explicit decoupling procedure. In \cite{sz}, the authors
discussed the mixing effect between metric fluctuation and Maxwell
fluctuation in RN-AdS spacetime. They  firstly derived the
diffusion constant through the mixed RG flow equations, and then
turn to the standard hydrodynamic calculation on the cut-off
surface by decoupling the equations of motion. It was found that
the  results through the RG flow approach  
match those through the hydrodynamic calculation in good
agreements.

In this  paper, we intend to apply these approaches to
investigate the vector-type fluctuations in
Einstein-Gauss-Bonnet-Maxwell theory. So far, the diffusion
coefficient for charged black holes in this theory has not been
computed in the previous literature. It is also interesting to
study the electric conductivity, charge  susceptibility and
thermal conductivity of  the dual fluid in this theory.
Unlike the RN-AdS spacetime, the equations of motion, followed the
method given in \cite{gmsst,gks},  are essentially failed to
be decoupled because of the mixing of the Gauss-Bonnet coupling
and the charge.  Nevertheless, in this paper we intend to
disclose one advantage of the membrane paradigm and the hWRG
approach,  which tell us that even the equations of motion
can not be decoupled, one can still calculate the transport
coefficients analytically. We will generalize the strategy
presented in \cite{strominger} to calculate the cutoff-dependent
diffusion constant $D(u_c)$ when the Gauss-Bonnet terms are
involved. We will also investigate the DC conductivity at an
arbitrary radius by using the Kubo formula without obtaining the
decoupled  matter equations as in \cite{sz}. The charge
susceptibility and the thermal conductivity are also evaluated. It
was suggested in \cite{kovtun} that the ratio of the conductivity
to the susceptibility may obey a universal bound. In \cite{myers},
it was shown that this bound is violated in the framework of
general four-derivative interactions. We will extend this
discussion in the Gauss-Bonnet gravity in this paper as well.

This paper is organized as follows. In section 2, we present our
 basic setup
for the vector type perturbation following the notation in
\cite{liu}. We define  the effective ``current'' and ``strength''
for the perturbation fields, and  then rewrite
the equations of motion in a form similar to the Maxwell equation.
In section 3, we first investigate the AC conductivity of the dual
plasma  with the use of the numerical  analysis. Then we
study the DC conductivity flow at an arbitrary radius  through
 the Kubo formula. Following the method developed in
\cite{faulk,strominger,sz}, we calculate the diffusion constant in
the scaling limit. The resulting formula for the diffusion
constant reproduces  the old result from the membrane
paradigm in the absence of the Gauss-Bonnet coupling. We then
provide a consistent check on the obtained diffusion constant by
 employing the Brown-York tensor on the cutoff surface.
Conclusion and discussion are presented in section 4. We also show
in the appendix the full equations of motions for vector type
perturbations.

\section{Basic setup}
We start by introducing the following action in $D$ dimensions
which includes Gauss-Bonnet terms and $U(1)$ gauge field:
\begin{equation}
\label{action}
I=\frac{1}{2\kappa^2}\!\int\!\dd^{D}\!x
\sqrt{-g}\Big(R-2\Lambda+\tilde{\alpha}\left(R_{\mu\nu\rho\sigma}
R^{\mu\nu\rho\sigma}-4R_{\mu\nu}R^{\mu\nu}+R^2\right)-\frac{4 \pi G_{D}}{g^2}
F_{\mu\nu}F^{\mu\nu}\Big),
\end{equation}
where $2\kappa^2=16 \pi G_{D}$, $g^2$ the $D$-dimensional gauge coupling constant,  and $\tilde{\alpha}$ is a
Gauss-Bonnet coupling constant with dimension $\rm(length)^2$.
The field strength is   defined  as $F_{\mu\nu}(x)=\del_\mu
A_\nu(x)-\del_\nu A_\mu (x)$.
The corresponding Einstein equation  reads as
\begin{eqnarray}
\label{einstein}
R_{\mu\nu}-\frac{1}{2}g_{\mu\nu}R+g_{\mu\nu}\Lambda
=\frac{8\pi G_{D}}{g^2}\Big(F_{\mu\rho}F_{\nu\sigma}g^{\rho\sigma}
-\frac{1}{4}g_{\mu\nu}F_{\rho\sigma}F^{\rho\sigma}\Big)
+T^{\rm eff}_{\mu\nu},
\end{eqnarray}
where
\begin{eqnarray}
T^{\rm eff}_{\mu\nu}
=\tilde{\alpha}
\Bigg[
&&
\frac{1}{2}g_{\mu\nu}
\Big(R_{\alpha\beta\rho\sigma}R^{\alpha\beta\rho\sigma}
-4R_{\alpha\beta}R^{\alpha\beta}
+R^2\Big)
-2RR_{\mu\nu}+4R_{\mu\rho}R_\nu{}^{\rho}
\nonumber\\
&&
+4R_{\rho\sigma}R_{\mu\nu}{}^{\rho\sigma}
-2R_{\mu\rho\sigma\gamma}R_{\nu}{}^{\rho\sigma\gamma}\Bigg].
\end{eqnarray}
The charged black brane/hole solution  to this equation in
$D$ dimensions
is described by~\cite{g2,g3,g4}
\begin{eqnarray}
\label{metric}
\dd s^2&=& -g_{tt}dt^2+g_{rr}dr^2+g_{ii}\dd x^{i}\dd x^{i}\nonumber\\
&=&
\displaystyle
-H(r)N^2\dd t^2+H^{-1}(r)\dd r^2+\frac{r^2}{l^2}
\dd x^{i}\dd x^{i},
\\
A_t
&=&
\displaystyle
\mu\bigg(1-\frac{r^{D-3}_{+}}{r^{D-3}}\bigg),
\end{eqnarray}

\vspace*{-6mm}
\noindent
with
\begin{eqnarray}
H(r)&=&k_{0}+\frac{r^2}{2\lambda
l^2}\left[1-\sqrt{1-4\lambda\bigg(1-\frac{r^{D-1}_{+}}{r^{D-1}}
-a\frac{r^{D-1}_{+}}{r^{D-1}}+a\frac{r^{2D-4}_{+}}{r^{2D-4}}\bigg)}\right],\\
\Lambda &=&-\frac{(D-1)(D-2)}{2l^2},
\end{eqnarray}
where  $\tilde{\alpha}$,  $\alpha$ and $\lambda$ are related by
$\alpha=(D-4)(D-3)\tilde{\alpha}$,  and $\lambda=\alpha/l^2$,
$a=\frac{l^2\kappa^2}{(D-2)(D-3)g^2}Q^2$  with the parameter
$l$ corresponding to AdS radius. The chemical potential $\mu$ is
related to $Q$ by $Q=\frac{\mu (D-3)}{r_{+}}$ and the horizon
$r_{+}$ is the largest root of $H(r)=0$. The charge density
evaluated at the asymptotic boundary is
$n_q=\frac{Qr^3_{+}}{g^2l^3}$.

The constant $N^2$ in the metric (\ref{metric}) can be fixed by
requiring that  the geometry of the spacetime should
asymptotically approach to the conformaly flat metric at spatial
infinity, i.e.\ $\dd s^2\propto -c^2\dd t^2+\dd\vec{x}^2$.
Since as $r\rightarrow\infty$, we have
$$
H(r)N^2 \rightarrow\frac{r^2}{l^2},
$$
one can find $N^2$ to be
\begin{equation}
N^2=\frac{1}{2}\Big(1+\sqrt{1-4 \lambda}\ \Big).\label{N}
\end{equation}
 Without loss of generality, in the following of this paper
we  will mainly focus on five-dimensional case $D=5$ with
$k_0 = 0$. After introducing the coordinate  as follows \be
u=\frac{r^2_{+}}{r^2},~~ H(u)=\frac{r^2_{+}}{ul^2}f(u),~~
f(u)=\frac{1}{2\lambda}\bigg(1-\sqrt{1-4\lambda(1-u)(1+u-au^2)}\bigg),
\ee the metric can  be rewritten as \be
ds^2=\frac{r^2_{+}}{l^2}\frac{1}{u}\bigg(-f(u)N^2dt^2+\sum^3_{i=1}dx^idx^i\bigg)+\frac{l^2}{4u^2}\frac{du^2}{f(u)}.
\ee The Hawking temperature  of the black brane can be worked out
as \be T_H=\frac{Nr_{+}}{2\pi l^2}(2-a). \ee We follow the
notation in \cite{liu} hereafter and note that  a rotational symmetry
 among
$x^i$  directions, i.e. $g_{ij}=g_{xx}\delta_{ij}$ with $x$
being one of the spatial direction.

The shear modes of the metric perturbations $h_{xz}$, $h_{xt}$,
$h_{xr}$
 exhibit the same behavior as the Maxwell field $a_z$, $a_t$ and
$a_r$, thus we can define \be a_z\equiv h^{x}_z, ~~~a_t\equiv
h^{x}_t,~~~a_u\equiv h^{x}_u. \ee Note that in the gauge $a_u=0$,
$a_t$ and $a_z$ decouple from the other components.   It turns
out that the vector part of the off shell action for the shear
modes in charged Gauss-Bonnet gravity is given by \be S=\int d^5 x
\sqrt{-g}\bigg[-\frac{1}{4g^2}F^{\mu\nu}F_{\mu\nu}-\frac{1}{4g^2_{eff}(r)}\bigg({f}^{ut}{f}_{ut}
+\tilde{f}^{uz}\tilde{f}_{uz}+\tilde{f}^{zt}\tilde{f}_{zt}\bigg)
+\frac{1}{g^2}\frac{4u^3}{r^2_{+}}a_t A'_xA'_t\bigg]. \ee
Let us define the  effective ``current''
and ``strength'' for the $a_\mu$ as \ba
&&j^{t}=\frac{1}{g^2_{eff}(r)}\sqrt{-g}\mathcal{M}uf^{ut},~~~
j^{z}=-\frac{1}{g^2_{eff}(r)}\sqrt{-g}\tilde{f}^{uz},\label{jz}\\
&&f_{ut}=\partial_u a_t-\partial_t a_u,\label{fut}\\~~~~~~
&&\tilde{f}_{uz}=-u^2\mathcal{M}'{f}_{uz}=-u^2\mathcal{M}'(\partial_u a_z-\partial_z a_u),\label{fuz}\\
&&\tilde{f}_{zt}=-u^2\mathcal{M}'{f}_{zt}=-u^2\mathcal{M}'(\partial_z
a_t-\partial_t a_z),\label{fzt} \ea where
$\mathcal{M}=\frac{1-2\lambda f}{u}$ and the prime $'$ denotes the
derivative with respect to $u$. $j^{\mu}$ is the conjugate
momentum of the field $a^{\mu}$. Because of the presence of the
Gauss-Bonnet terms, the  definition of the effective current
and strength becomes a little bit complicated than that of the Einstein
case\cite{sz} and it can return to the  Einstein
case without the
Gauss-Bonnet terms.  For chargeless Gauss-Bonnet black branes, the equations of
motion can be written as \ba
&&\partial_tj^t+\partial_zj^z=0,\\
&&\partial_uj^t+Gg^{tt}g^{zz}\partial_z\tilde{f}_{zt}=0,\\
&&\partial_uj^z-Gg^{tt}g^{zz}\partial_z\tilde{f}_{zt}=0, \ea where
we have used the notations \be
G=\sqrt{-g}/g^2_{eff}(r)=\frac{r^6_{+}N}{4\kappa^2
u^4l^5},~~~\frac{1}{g^2_{eff}(r)}=\frac{r^2_{+}}{2\kappa^2 l^2 u}.
\ee For the charged black brane there is a matter perturbation
$A_x$, and in this case we  may define the charge density as \be
\tilde{j}^t={j}^t-\frac{1}{g^2}\sqrt{-g}\frac{4u^3}{r^2_{+}}A'_tA_x.
\ee The equations of motion then take the following form \ba
\partial_t\tilde{j}^t+\partial_zj^z&=&0,\label{m1}\\
\partial_u\tilde{j}^t+Gg^{tt}g^{zz}\partial_z\tilde{f}_{zt}&=&0,\label{m2}\\
\partial_uj^z-Gg^{tt}g^{zz}\partial_t\tilde{f}_{zt}&=&0.\label{m3}
\ea
The Bianchi identity yields
\be
\partial_u\bigg( \frac{\tilde{f}_{zt}}{-u^2\mathcal{M}' }\bigg)-\frac{g_{uu}g_{tt}\partial_z j^t}{u\mathcal{M} G}+\frac{\partial_t j_z}{u^2\mathcal{M}'G}=0.
\ee
The equation of motion for $A_x$ can be written as
\be
\partial_u \tilde{J}^x-\frac{1}{g^2 N^2}g^{xx}\bigg(-g^{tt}N^2\partial_t F_{tx}+g^{zz}\partial_z F_{zx}\bigg)=0,
\ee where the current $\tilde{J}^x$ containing the mixed term  is
defined as \be
\tilde{J}^x=J^x+\frac{1}{g^2}\sqrt{-g}\frac{4u^3}{r^2_{+}N^2}A'_t
a_t,~~~J^x=-\frac{1}{g^2}\sqrt{-g}g^{uu}g^{xx}\partial_u A_x. \ee
{Here $F_{\mu\nu}$ characterize the strength of} the Maxwell
fields  $A_\mu$ and should not be confused with the effective
strength  of the vector modes of gravity $f_{\mu\nu}$. The
matter perturbation $A_x$ is also constrained by the Bianchi
identity \ba
\partial_u F_{xt}-\frac{g^2 g_{uu}g_{xx}}{\sqrt{-g}}\partial_t J^x=0,\\
\partial_z F_{tx}+\partial_t F_{xz}=0.
\ea
\section{Transport coefficients}
In this section, we will evaluate
  the transport coefficients of the dual plasma on the cutoff
 surface by solving the equations of motion for charged black
holes.  We will first compute the AC conductivity
numerically and then derive the DC conductivity by using the Kubo formula.  We derive the cutoff
dependent diffusion constant in the scaling limit. Finally, we provide a consistent check on the obtained diffusion constant.

\subsection{AC electric conductivity flow at $k_z=0$}
 We will
work in Fourier space. For  instance, we  expand $a_z$ as \be
a_z(t,z,u)=\int \frac{d^4k}{(2\pi)^4}e^{-i\omega t+ik z}a_z(k,u).
\ee The same decomposition is  applicable to  $a_t$ and
$A_x$.
In the zero momentum limit, $A_x$ decouples from $a_t$
\be\label{ax}
\partial_u\bigg(\sqrt{-g}g^{uu}g^{xx}\partial_uA_x\bigg)
-\frac{g^2_{eff}}{g^2}\sqrt{-g}(A'_t)^2\frac{4u^2}{\mathcal{M}r^2_{+}}A_x+\omega^2\sqrt{-g}g^{xx}g^{tt}A_x=0.
\ee The on-shell action for $A_x$ at the cut-off surface $u=u_c$
is given by \be S=\int \tilde{J}^x A_x. \ee The electric
conductivity  of the plasma on this cut-off surface can be
defined as \be \sigma_A(\omega, u_c)=\frac{J^x}{i\omega A_x}. \ee
Then equation (\ref{ax}) can be recast as \be\label{num}
\frac{\partial_{u_c}\sigma_A}{i\omega}-\frac{g^2\sigma^2_A}{\sqrt{-g}g^{uu}g^{xx}}-\frac{g^2_{eff}
\sqrt{-g}}{g^4\omega^2}(A'_t)^2\frac{4u^2}{\mathcal{M}r^2_{+}}+\frac{1}{g^2}\sqrt{-g}g^{xx}g^{tt}=0.
\ee  The regularity condition on the horizon gives
\be\label{SA} \sigma_A(u=1)=\frac{1}{g^2}\frac{r_{+}}{l}, \ee in
 agreement with the result given in \cite{liu}. It is worth
noting that the conductivity at the event horizon neither depends
on the charge, nor the Gauss-Bonnet coupling. In other words, we
can say the regularity at the horizon corresponding to setting \be\label{bc}
J^x(1)=i\omega \sigma_A(1) A_x(1)=i\omega \frac{1}{g^2
}\sqrt{\frac{-g}{g_{uu}g_{tt}}}g^{xx}\bigg|_{u=1} A_x(1). \ee This
relation is useful in determining the DC conductivity.

In order to have  an explicit picture on the solutions of
conductivity, we solve equation (\ref{num}) numerically and plot
the conductivity flows with different charges in Figure 1. From
this figure, we can see that when the charged black brane becomes extremal, there is a fixed point for the flow
of conductivity near the horizon due to the appearance of $\rm AdS_2$ near the horizon. Same as \cite{sz}, this fixed point will disappear in non-extremal case. The numerical calculation also
implies that as $\omega$ approaches  to zero, the real part
of the conductivity behaves more like equation (\ref{dc}).  The
numerical results in Figure 1  change little for different values
of Gauss-Bonnet coupling.
\begin{figure}[htbp]
 \begin{minipage}{1.1\hsize}
\begin{center}
\includegraphics*[scale=0.4] {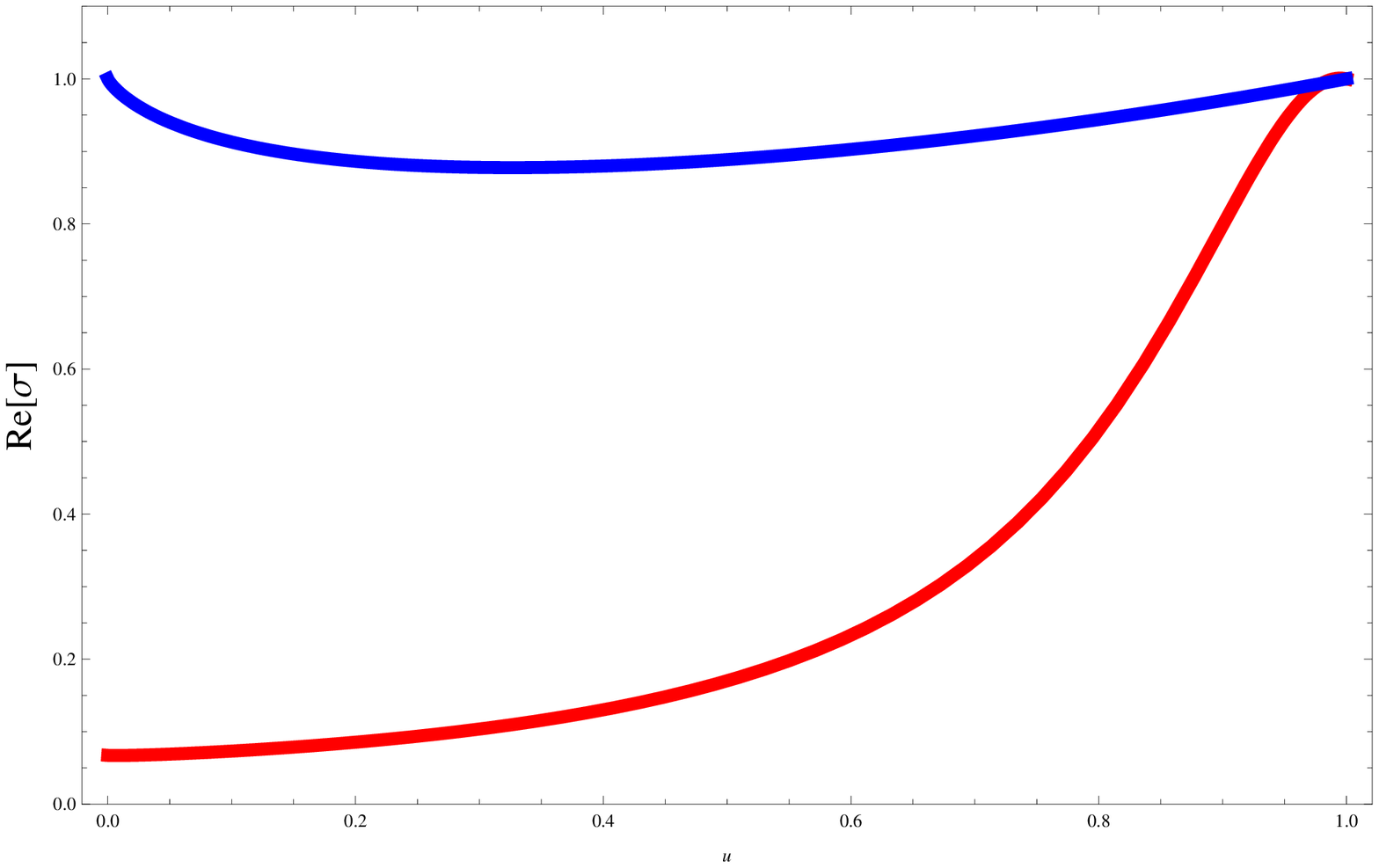}
\includegraphics*[scale=0.4]{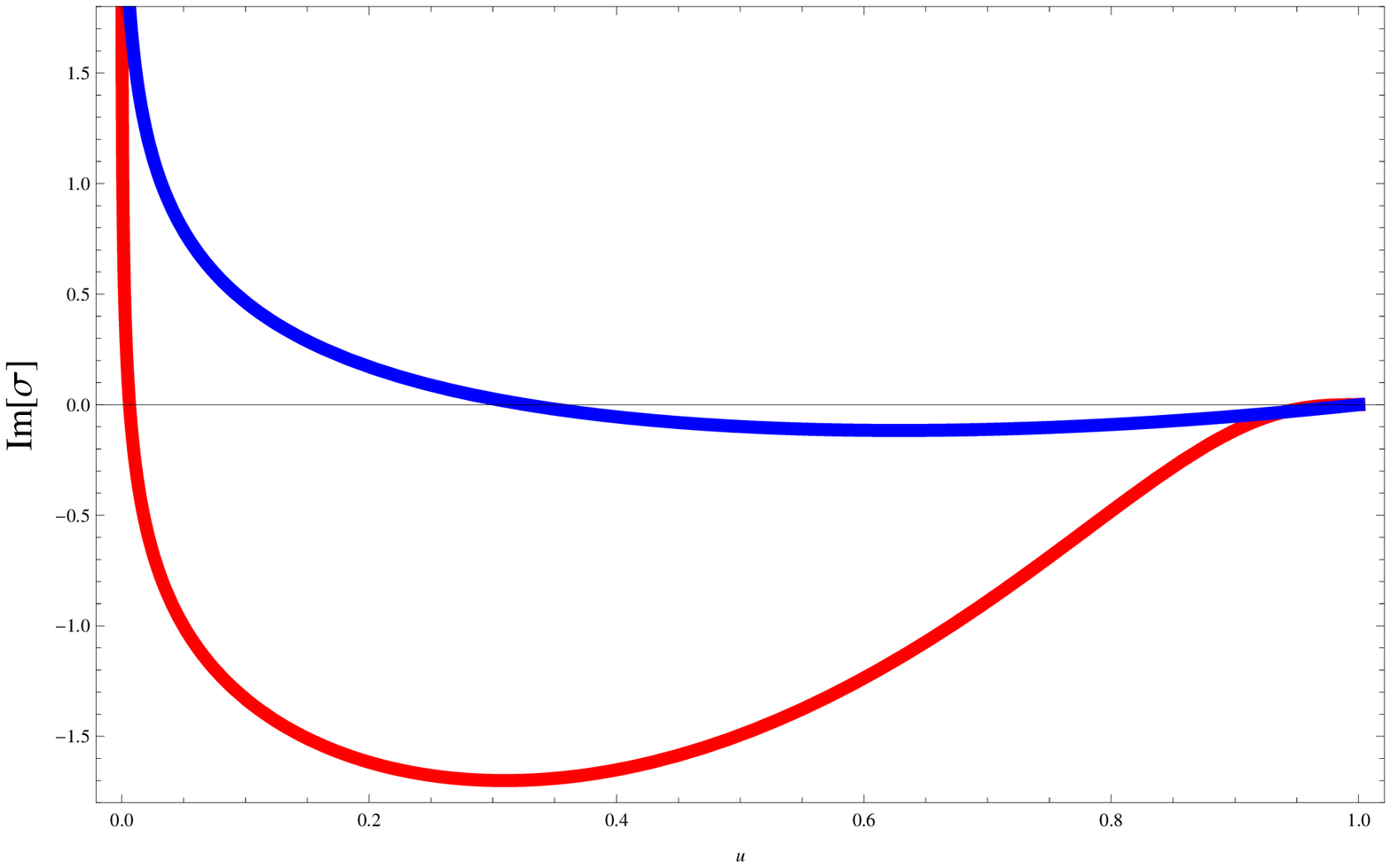}
\end{center}
\caption{ The radial flow of AC conductivity in five-dimensional
RN-AdS-Gauss-Bonnet black holes for non-extremal case (blue line)  and
extremal case (red line) from up to bottom. Here we set $\lambda=0.01$, $r_{+}=1$, $\mu=0.1$, $g=l=G_5=1$ and $\omega=0.978$ for the non-extremal case, but $a=2$ for the extremal case.  } \label{re}
\end{minipage}
\end{figure}
\subsection{DC conductivity flow}
We now turn to   the DC conductivity in the perturbative
background. The conductivity can be evaluated by using  the Kubo
formula, which relates the conductivity to the low frequency and
zero momentum limit of the retarded Green's function \cite{pss,
ss} \be G^{R}_{x,x}(\omega,k=0)=-i\int dt d\textbf{x}e^{i\omega
t}\theta(t)<[J_x(x),J_x(0)]>, \ee where $J_{\mu}$ is the conformal field theory (CFT)
current dual to the bulk gauge field $A_{\mu}$. The DC
conductivity is given by \be \sigma_{DC}=-\lim_{\omega\rightarrow
0}\frac{e^2{\rm Im} G^{R}_{x,x}(\omega,k=0)}{\omega}. \ee
According to \cite{kovtun}, the constant $e$ marks the coupling
between the boundary current and an external or auxiliary vector field.
To the leading order of $e$, the effects of the auxiliary vector
are negligible and the conductivity can be fixed from the original
CFT. In the absence of chemical potential the general form of
conductivity was derived in \cite{liu}  through the membrane
paradigm. But when the chemical potential is  taken into
account, those formulae need revision as there is a non-trivial
flow from horizon to boundary. At the linear level of equations
and in the zero momentum limit metric perturbation $a_z$ decouples
from the rest (see the Appendix for more details). The
corresponding component of Einstein's equations becomes a
constraint \be\label{at}
 {a_t}'=\frac{3aN^2B(u)}{\mathcal{M}},
\ee where $B$ is defined as $B=\frac{A_x}{\mu}=\frac{2A_x}{Qr_{+}}$. Equation
(\ref{at}) can also be obtained in the scaling limit $\omega \sim k^2
\ll 1$.  After substitution one finds the  equation for perturbed
gauge fields to be \be\label{b}
B''+\frac{f'}{f}B'+\frac{l^4b^2}{4N^2uf^2}\bigg(\omega^2-k^2f\bigg)B-\frac{3a}{\mathcal{M}f}B=0.
\ee The effective action up to the quadratic order takes the
simple form \be
I^{(2)}=\frac{1}{2\kappa^2}\int\frac{d^4k}{(2\pi)^4}du\bigg(K(u)B'(u,k)
B'(u,-k)+L(u)B(u,k) B(u,-k)\bigg), \ee where \be
K(u)=-\frac{\sqrt{-g}}{g^2}g^{xx}g^{uu}\mu^2,~~~L(u)=\bigg(\omega^2\frac{\sqrt{-g}}{g^2}g^{xx}g^{tt}-\frac{g^2_{eff}}{g^4}(A'_t)^2\frac{4u^2}{\mathcal{M}r^2_{+}}\bigg)\mu^2.
\ee At this point, it is convenient to define the radial momentum
as \be J^x_k\equiv \frac{\delta I^{(2)} }{\delta B'(u,k)
}=\frac{1}{\kappa^2}K(u)B'(k). \ee The equation of motion for $B$
then takes the form \be
\partial_uJ^x_k=\frac{1}{\kappa^2}L(u)B(u,k).
\ee The regularity at the horizon $u=1$ corresponds to \cite{liu}
\be J^x_k(1)=-i\omega {K(u)}\sqrt{\frac{g_{uu}}{g_{tt}}}B(1). \ee
The same boundary condition has been obtained in equation (\ref{bc}).
Following the standard procedure for the computation of retarded
Green's functions in Minkowski spacetime \cite{pss,ss}, we now evaluate the
effective action on-shell and obtain the boundary term, which is
given by \be
I^{(2)}=\int\frac{d^4k}{(2\pi)^4}\mathcal{F}_k\bigg|^{u=1}_{u=u_c}.
\ee In order to  evaluate the conductivity at a finite cutoff
surface, we need remove the limit $u\rightarrow 0$ away from the
UV cutoff. Note that the expression is evaluated on
$B(u,\omega,k)$ with infalling boundary conditions at the horizon.
In the present case, one  evaluates the on-shell action to write
down the flux factor as \be 2\mathcal{F}_k=J^x_k B(u,-k)
\ee For simplicity, 
we set $e=1$ and $\kappa^2=1$ hereafter. The DC conductivity at a
finite cutoff surface is then given by \be
\sigma_{DC}=\lim_{\omega\rightarrow 0,u\rightarrow u_c} {\rm
Im}\bigg(\frac{2\mathcal{F}_k}{A_x(u)
A_x(u)}\bigg)\bigg|_{k=0}=\lim_{\omega\rightarrow 0, u\rightarrow
u_c}\frac{{\rm Im}[ J^x_k B(u,-k)]}{\omega \mu^2
B(u,k)B(u,-k)}\bigg|_{k=0}. \ee Keeping in mind that $J^x_k$ is
constrained by the regularity condition at the horizon, we may
write \be
\sigma_{DC}=-\frac{K(u)}{\mu^2}\sqrt{\frac{g_{uu}}{g_{tt}}}
~~\bigg|_{u=1}\frac{B(1)B(1)}{B(u_c)B(u_c)}. \ee Now we need to
solve $B(u)$ by imposing  the regularity  condition at the horizon
and setting $\omega$ to zero. We can easily solve equation
(\ref{b}) in the case $\omega=0$ and $k=0$: \be
B(u)=B(0)\frac{2(1+a)-3a u}{2(1+a)}. \ee Note that it is not
proper to impose the boundary condition that $B(u)$ must be
vanishing at the cutoff surface because the DC conductivity
would be divergent under such boundary condition. Finally, we
obtain the DC conductivity at the cutoff surface\be\label{dc}
\sigma_{DC}=\frac{r_{+}}{g^2 l}\frac{(2-a)^2}{[2(1+a)-3au_c]^2}.
\ee This result is consistent with the numerical calculation
conducted in section 3 in the small $\omega$ limit (see Figure 2
for details).
\begin{figure}[htbp]
 \begin{minipage}{1.1\hsize}
\begin{center}
\includegraphics*[scale=0.6] {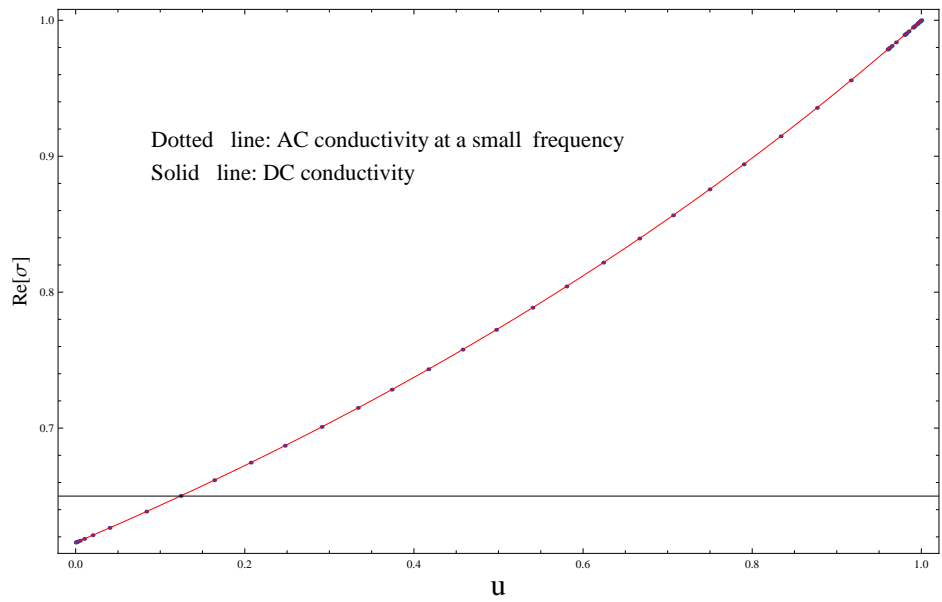}
\end{center}
\caption{ Flows of DC conductivity and numerical result for AC
conductivity at a small frequency $\omega=0.01$. These two results
matches with each other. Here we set $\lambda=0.01$, $r_{+}=1$, $\mu=0.1$ and $g=l=G_5=1$. } \label{re}
\end{minipage}
\end{figure}
At the horizon $u_c=1$, the above equation becomes \be \sigma_{\rm
DC}=\frac{r_{+}}{g^2 l}, \ee which is consistent with (\ref{SA}).
As one goes to the boundary $u_c\rightarrow 0$, the DC
conductivity is reduced to \be \sigma_{\rm DC}=\frac{r_{+}}{g^2
l}\frac{(2-a)^2}{4(1+a)^2}. \ee which agrees with  the
previous result in \cite{gmsst}. In the absence of the chemical
potential, (\ref{dc}) is reduced to the formula for DC
conductivity given in \cite{liu}. However, in \cite{liu} it was
assumed that the background configuration of the gauge field is
not perturbed.  As a matter of fact, the fluctuations of the
metric and the gauge field are coupled such that the contribution
of the chemical potential should not be neglected.

 Next we turn to discuss the ratio of conductivity to the
susceptibility. For simplicity, in the following calculation we
set all the coupling parameters to be  unit (i.e.
$\kappa^2=e=g^2=1$) and also $l=1$.  In conformal field
theory, the ratio of the conductivity to the susceptibility does
not depend on the relative normalization of the current and the
stress tensor.  Since the susceptibility is defined as \be
\chi\equiv \frac{\partial n_q}{\partial \mu}\bigg|_T, \ee where
$n_q$ is the charge density at the cut-off surface,  then the
expression for $\frac{\sigma_{DC}}{\chi}$  can be written as \be
\frac{\sigma_{DC}}{\chi}=\frac{1}{16\pi
T}\bigg(\frac{2-a}{1+a}\bigg)^2=\frac{1}{4\pi
T}(1-3a)+\mathcal{O}(a^2). \ee In  the absence of the chemical
potential $\mu=0$ (i.e. $a=0$), the above equation recovers the
Einstein relation $\frac{\sigma_{DC}}{\chi}\bigg|_{\mu=0}=D_c$ with
$D_c=\frac{1}{4\pi T}$, denoting the  diffusion constant for
 Schwarzschild-AdS
 black holes \cite{strominger,gmsst}.  Since both $\sigma_{DC}$ and $\chi$ do not depend on the Gauss-Bonnet coupling, the above result implies that the Einstein relation
 cannot be used to derive the diffusion constant in the presence of higher-derivative gravity corrections \cite{myers}.  After obtaining the
conductivity, the thermal conductivity $\kappa_T$  can be computed
 using the formula \cite{Rcharged} \be
\kappa_T=\bigg(\frac{\epsilon+P}{T_H n_q}\bigg)^2 {T_H}
\sigma_{DC}=\frac{(2-a)\pi r^2_{+}}{3a}. \ee The ratio
$\frac{\kappa_T \mu^2}{\eta T_H}$  was computed in \cite{Rcharged}
\be \frac{\kappa_T \mu^2}{\eta
T_H}=\frac{2\pi^2}{1-2\lambda(2-a)}=2\pi^2\bigg[1+2(2-a)\lambda+4(2-a)^2\lambda^2+...\bigg],
\ee where we have used (\ref{shear}). One may find that the
Gauss-Bonnet coupling modify the leading behavior. Note that the
leading order result given in \cite{Rcharged} was $8\pi^2$ because
the normalization for the gauge kinetic term differs by a factor
of $4$ from that used here.

As a conjecture, we propose that the DC conductivity for single
charged black branes can be evaluated  through the thermal
quantities (see also \cite{hartnoll}) \be\label{con} \sigma_{\rm
DC}=\frac{1}{g^2
}\sqrt{\frac{-g}{g_{uu}g_{tt}}}g^{zz}\bigg|_{u=1}\bigg(\frac{sT_c}{\varepsilon+p}\bigg)^2
=\frac{1}{g^2}\sqrt{\frac{-g}{g_{uu}g_{tt}}}g^{zz}\bigg|_{u=1}\bigg(1-\frac{n_q\mu_c}{\varepsilon+p}\bigg)^2,
\ee where $s=\frac{1}{4G}u^{\frac{3}{2}}_c$ denotes the entropy
density on the cut-off surface. Note that we have used the
thermodynamical relation \be \varepsilon+p-sT_c-n_q\mu_c=0. \ee
Here the charge density  and chemical potential are $n_q=Q
u^{3/2}_c$ and $\mu=\frac{ l Q u^{1/2}_c}{2
\sqrt{f(u_c)}}(1-u_c)$, respectively. One may find that
(\ref{con}) can recover the DC conductivity obtained in
\cite{sz,gks} for four-dimensional RN-AdS black holes.  The
expression (\ref{con})  is also consistent with
\cite{jain} in the single charge case.

\subsection{Diffusion constant $\bar{D}(u_c)$ at the cutoff surface}

Now we evaluate the ``conductivity'' introduced by the metric
perturbation $a_z$ at zero momentum. The conductivity in this case
can be defined as \be \sigma_h:=\frac{j^z}{f_{zt}}=(-u^2
\mathcal{M}' )\frac{j^z}{\tilde{f}_{zt}}. \ee In the zero momentum
limit,  the decoupled flow equation for $\sigma_h$ is given
by \be \frac{\partial_{u_c}\sigma_h}{-i\omega}+\frac{1}{(-u^2
\mathcal{M}' )}\sigma^2_h\frac{g_{uu}g_{zz}}{G}-(-u^2 \mathcal{M}'
)G g^{tt}g^{zz}=0. \ee Again  the regularity  condition
at the event horizon gives \be\label{shear}
\sigma_h(u=1)=\frac{1}{16 \pi
G_N}\bigg(\frac{r^3_{+}}{l^3}\bigg)\bigg[1-2\lambda(2-a)\bigg].
\ee This is actually the  shear viscosity for charged
Gauss-Bonnet-AdS black branes because when the momentum is
vanishing and thus no polarization direction, the conductivity is
exactly the shear viscosity. It is worth noting that (\ref{shear})
indicates that the conductivity at the horizon obeys \be\label{bb}
j^z(1)=\sigma_h(1)f_{zt}(1). \ee This relation will be used in the
following computation.\\

$\bullet$ \textsl{Diffusion constant}\\

In the following, we treat the vector modes in a long wave-length
expansion for fields and the equations of motion. The diffusion constant will be evaluated by a line
integral from the horizon to the cutoff surface. In
\cite{strominger}, it was found that the diffusion constant runs
 with the variation of $u_c$. We will show as below that the
diffusion constant not only runs  with $u_c$, but also
depends on the Gauss-Bonnet coupling constant. For non-vanishing
momentum $k_z$, the equation of motion for $\sigma_h$ is coupled
to other modes. To proceed further we take  the scaling limit
for temporal and spatial derivatives as  \ba
\partial_t \sim \epsilon^2,~~~~ \partial_z \sim \epsilon,\\
\tilde{f}_{zt}\sim \epsilon^3\bigg(\tilde{f}^{(0)}_{zt}+\epsilon
\tilde{f}^{(1)}_{zt}+... \bigg), \ea The in-falling boundary
condition at the horizon requires $j^z$ is linearly related to
$\tilde{f}_{zt}$. Therefore, we have \be j^z\sim
\epsilon^3\bigg(j^{z(0)}+\epsilon j^{z(1)}+...\bigg). \ee From the
charge conservation equation, we find that \ba
\tilde{j}^t \sim \epsilon^2 \bigg(\tilde{j}^{t(0)}+\epsilon \tilde{j}^{t(1)}+...\bigg),\\
j^t\sim \epsilon^2 \bigg({j}^{t(0)}+\epsilon j^{t(1)}+...\bigg).
\ea In the lowest order, (\ref{m2}) and (\ref{m3}) reduce to \be
\partial_{u}\tilde{j}^{t(0)}=0,~~~\partial_{u}j^{z}=0.
\ee The above first equation indicates that \be\label{tjt}
\tilde{j}^{t(0)}={j}^{t(0)}-\frac{1}{g^2}\sqrt{-g}\frac{4u^3}{r^2_{+}}A'_tA^{(0)}_x=C_0,
\ee  and $C_0$ is a constant. The charge conservation
equation (\ref{m1}) requires \be
j^{z(0)}=\tilde{j}^{t(0)}\frac{\omega}{k}. \ee The lowest order
 of the Bianchi identity becomes \be
\partial_u \bigg(\frac{\tilde{f}^{(0)}_{zt}}{-u^2 \mathcal{M}'}\bigg)=\frac{g_{uu}g_{tt}}{u \mathcal{M}G}\partial_z {j}^{t(0)}.
\ee
Evaluated at the horizon $u=1$, the above equation has the solution
\be\label{tfzt}
 \frac{\tilde{f}^{(0)}_{zt}(u)}{u^2 \mathcal{M}'}= \frac{\tilde{f}^{(0)}_{zt}}{u^2 \mathcal{M}'}\bigg|_{u=1}+2\kappa^2\int^1_{u}du\frac{\partial_z {j}^{t(0)}}{\sqrt{-g}g^{uu}g^{tt}g_{xx}u \mathcal{M}}
\ee Using the definition of the current given in (\ref{fzt}), the
above equation can be written as \be\label{lf}
f_{zt}(u)=-\frac{\tilde{f}^{(0)}(1)}{1-2(2-a)\lambda}+2\kappa^2\int^1_{u}du\frac{\partial_z
{j}^{t(0)}}{\sqrt{-g}g^{uu}g^{tt}g_{xx}u \mathcal{M}} . \ee In the
scaling limit, the Maxwell equation for the gauge perturbation
$A^{(0)}_x$  becomes \be\label{Ax}
\partial_u\bigg(\sqrt{-g}g^{uu}g^{xx}\partial_u A^{(0)}_x\bigg)=\sqrt{-g}\frac{4u^3}{r^2_{+}}A'_t {a}'^{(0)}_t.
\ee Before  evaluating the diffusion constant $D_c$ with the
use of (\ref{lf}), we should first find  the solution for
$A^{(0)}_x$. Following \cite{strominger},  we substitute
(\ref{Ax}) into (\ref{tjt}), further impose the boundary condition
$A^{(0)}_x(u_c)=0$, and then we obtain \be
A^{(0)}_x(u)=\frac{3a}{8\pi G
[2(1+a)-3a u_c]Q}\bigg(1-\frac{u}{u_c}\bigg). \ee
 Consequently, we have \be\label{lf1}
f_{zt}(u)=\frac{\tilde{f}^{(0)}(1)}{1-2(2-a)\lambda}+\frac{i k
f(u_c)l^2}{2 r_{+}[2(1+a)-3a u_c]}. \ee Following the sliding
membrane paradigm \cite{liu}, we define the conductivity by
the current and electric fields as \be \sigma_h
(u_c):=\frac{j^{z(0)}(u_c)}{f^{(0)}_{zt}(u_c)}=\frac{\omega}{k}\frac{\tilde{j}^{t(0)}(u_c)}{f^{(0)}_{zt}(u_c)}.
\ee By further using the boundary condition given in (\ref{bb}), we find the expression for the conductivity $\sigma_h$ from
(\ref{lf1}) \be\label{cond}
\frac{1}{\sigma_h(u_c)}=\frac{1}{\sigma_h(1)}-\frac{k^2}{i\omega}\frac{{D}(u_c)}{\sigma_h(1)},
\ee where the diffusivity is \be
D(u_c)=\frac{\frac{l^2}{2r_{+}}f(u_c)}{2(1+a)-3a u_c}\bigg[1-2(2-a)\lambda\bigg].
\ee
The value of $D(u_c)$ is dimensional and can be set to any value through a coordinate transformation. It is meaningful to define a dimensionless diffusion constant as what follows.\\

$\bullet$ \textsl{Normalization of the diffusion constant}\\

We can introduce the proper frequency $\omega_c$ and  the
proper momentum $k_c$ conjugate to  the  proper time and
the proper  distance  respectively on the hypersurface
$u=u_c$. We define \be \omega_c\equiv
\frac{\omega}{\sqrt{g_{tt}}},~~~~~~~k_c\equiv
\frac{k}{\sqrt{g_{ii}}}. \ee The Hawking temperature at the
cut-off surface $u=u_c$ is determined by the Tolman relation \be
T_c(u_c)=\frac{T_H}{\sqrt{g_{tt}}},
~~~~T_H=\frac{f'(r_{+})}{4\pi}. \ee The coordinate-invariant,
dimensionless diffusion constant $\bar{D}(u_c)$ can be defined
as\cite{strominger} \be \bar{D}(u_c)=D(u_c)T_c
\frac{g_{zz}}{\sqrt{g_{tt}}}. \ee Thus the conductivity can
be written in terms of the normalized momentum and the diffusion
constant as follows \be\label{cond}
\frac{1}{\sigma_h(u_c)}=\frac{1}{\sigma_h(1)}-\frac{k^2_c}{i\omega_c}\frac{\bar{D}(u_c)/T_c}{\sigma_h(1)},
\ee where the dimensionless diffusion constant is given by
\be\label{bd}
\bar{D}(u_c)=\frac{1}{4\pi N}\frac{2-a}{2(1+a)-3au_c}\bigg[1-2\lambda(2-a)\bigg].
\ee The diffusion constant obtained here depends on the charge,
the position of the cut-off surface and the Gauss-Bonnet coupling.
Since the conductivity and the retarded Green function has the
relation \be \sigma^{ij}(k_\mu, u_c)=\frac{G^{ij}_R(k_\mu,
u_c)}{i\omega}. \ee From (\ref{cond}), we can write the related
Green function $G_{z,z}$ as \be G_{z,z}=\frac{\omega^2_c
\sigma_h(1)}{i\omega_c-\bar{D}(u_c)k^2/T_c}, \ee where the
expression for $\sigma_h(1)$ is given by \be
\sigma_h(u=1)=\frac{1}{16 \pi
G_N}\bigg(\frac{r^3_{+}}{l^3}\bigg)\bigg[1-2\lambda(2-a)\bigg].
\ee  Our results obtained above can recover equation (4.17c)
of reference \cite{gmsst}, when $\lambda\rightarrow 0$ and
$u_c\rightarrow 0$.

\subsection{A consistent check on $\bar{D}(u_c)$}
In this section, we provide a consistent check on the diffusion
constant by  employing the Brown-York stress tensor $t_{ij}$
on the cut-off surface. Since the detailed computation has been
presented in \cite{niu}, we only summarize the main results here.
The induced metric on the cutoff surface $u=u_c$ outside the
horizon is \be
ds^2_c=\frac{r^2_{+}}{l^2}\frac{1}{u_c}\bigg(-f(u_c)N^2dt^2+\sum^3_{i=1}dx^idx^i\bigg)
\ee
The Brown-York stress tensor on the hypersurface 
with unit normal is defined by \be
t_{ij}=\frac{1}{2\kappa^2}\bigg[K g_{ij}-K_{ij}-2\lambda(3
J_{ij}-J g_{ij})-\mathcal{C}g_{ij}\bigg], \ee with \be
J_{ij}=\frac{1}{3}\bigg(2K
K_{ik}K^k_j+K_{kl}K^{kl}K_{ij}-2K_{ik}K^{kl}K_{lj}-K^2K_{ij}\bigg),
\ee where  $K_{ij}$ is the extrinsic curvature, $K\equiv g^{ij}K_{ij}$ and  $\mathcal{C}$ is a constant.
 On the other hand, the stress-energy tensor of a fluid in
equilibrium has its form \be t_{ij}=(\epsilon+p)u_i u_j+pg_{ij},
\ee with $\epsilon$ the energy density, $p$ the pressure and $u^i$
the normalized fluid four-velocity. It was shown in \cite{niu}
that the shear viscosity from the Brown-York tensor is given by
\be \eta=\frac{1}{16\pi G l^3}u^{\frac{3}{2}}_c
\bigg[1-2\lambda(2-a)\bigg] \ee The sum of the energy density
$\varepsilon$ and pressure $p$ is found to be \cite{niu} \be
\varepsilon+p=\frac{\bigg[2(1+a)u^2_c-3a u^3_c\bigg]}{8\pi G
l\sqrt{f(u_c)}}. \ee It is easy to verify that \be
\bar{D}(u_c)=\frac{\eta}{\varepsilon+p}T_c=\frac{1}{4\pi N}\frac{2-a}{2(1+a)-3au_c}\bigg[1-2\lambda(2-a)\bigg],
\ee which is consistent with (\ref{bd}).

\section{Conclusion and discussion}

In this paper, we have investigated the RG flows for transport
coefficients for quark gluon plasma at finite chemical potential
with charged  AdS-Gauss-Bonnet black hole dual. We write down
the mixed flow equations explicitly and study the corresponding
conductivities. We  derive an expression for the diffusion
constant at a finite cutoff surface in the presence of
non-vanishing chemical potential and the Gauss-Bonnet terms. The
effect of turning on chemical potential can be thought of as
turning on effective interaction. Actually, the mixing effect of
metric and Maxwell fluctuations in charged black hole is very
important because the mixing effect tells us how transverse vector
modes of Maxwell fields can diffuse and longitudinal Maxwell modes
can have sound modes in the presence of chemical potential.

While the diffusion constant is sensitive to the Gauss-Bonnet
coupling, the computation of the DC conductivity seems to be
independent of the  Gauss-Bonnet coupling numerically and
analytically. In addition, the structure of this paper is
different from \cite{sz} since the DC conductivity is not derived
from the decoupled master equations.  We show that in the long
wavelength limit, we only need to solve equations up to zeroth
order in $\omega$.

\vspace*{10mm} \noindent
 {\large{\bf Acknowledgments}}\\
We would like to thank the KITPC for hospitality during the course
of the programm ``String Phenomenology and Cosmology '' when this
work is completed. The work of XHG  was partly supported by NSFC,
China (No. 11005072),  Shanghai Rising-Star Program and Shanghai
Leading Academic Discipline Project (S30105). YL was partly
supported by NSFC (10875057), Fok Ying Tung Education Foundation
(No.111008), the key project of Chinese Ministry of Education
(No.208072), Jiangxi young scientists (JingGang Star) program and
555 talent project of Jiangxi Province. YT and XW was partly
supported by NSFC (Nos.
10705048, 10731080 and 11075206) and the President Fund of GUCAS.\\
\appendix
\section{ Equations of motion for vector type fluctuations}
We consider the shear mode in the five-dimensional
RN-AdS-Gauss-Bonnet background by choosing the following gauge $
a_r=0,~A_r=0, $ and use the Fourier decomposition \ba
a_\mu(t,z,u)=\int \frac{d^4k}{(2\pi)^4}e^{-i\omega t+ik z}a_\mu(k,u),\\
A_x(t,z,u)=\int \frac{d^4k}{(2\pi)^4}e^{-i\omega t+ik z}A_x(k,u),
\ea
where we choose the momenta which are along the $z-$direction.
Note that after introducing $\mathcal{M}=\frac{1-2\lambda f}{u}$, the equations of motion can be written as
\ba
\label{21}&&0={a_t}''+\frac{\mathcal{M}'}{\mathcal{M}}{a_t}'+\frac{l^4b^2}{4f(u)}\frac{\mathcal{M}'}{\mathcal{M}}(k^2a_t+\omega k a_z)-\frac{3aN^2B'(u)}{\mathcal{M}},\\
\label{22}&&0=\omega {a_t}'-\frac{u\mathcal{M}'}{\mathcal{M}}f(u)N^2k {a_z}'-\frac{3aN^2B(u)\omega}{\mathcal{M}} ,\\
\label{23}&&0={a_z}''-\frac{\frac{1}{u}-\frac{f'(u)}{f}+2\lambda[\frac{f(u)}{u}+u f''(u)+uf'(u)^2/f(u)-2f'(u)]}{1+2\lambda (uf'(u)-f(u))}{a_z}'\nonumber\\&&+\frac{l^4b^2}{4N^2uf^2(u)}(\omega k a_t+\omega^2 a_z),\\
\label{24}&&0=B''+\frac{f'}{f}B'+\frac{l^4b^2}{4N^2 uf^2}(\omega^2- k^2 f)B-\frac{1}{N^2f}{a_t}',
\ea
where $b=\frac{l^2}{r_{+}}$, $g=1$ and the field $B$ is defined as $B=\frac{A_x}{\mu}=\frac{2A_x}{Qr_{+}}$. It is almost impossible  to decouple the above equations of motion because of the mixing of the Gauss-Bonnet coupling and  the charge. We  leave this to the future work.  For neutral black holes in Gauss-Bonnet gravity, the shear modes can be decoupled and it was analyzed in great detail by using the Kubo formula \cite{bri,kats,buchel}. For tensor type perturbation for charged black holes in Gauss-Bonnet gravity, one may refer to \cite{ge,cai} (see also \cite{boe,ce,hu,pang, af, neu}).
\renewcommand{\theequation}{A.\arabic{equation}}

\setcounter{equation}{0} \setcounter{footnote}{0}

\end{document}